\documentstyle[aps,preprint,floats,epsf,epsfig]{revtex}
\begin{document}
\def\be{\begin{eqnarray}}
\def\en{\end{eqnarray}}
\def\non{\nonumber}
\def\la{\langle}
\def\ra{\rangle}
\def\nc{N_c^{\rm eff}}
\def\vp{\varepsilon}
\def\vma{{_{V-A}}}
\def\vpa{{_{V+A}}}
\def\m{\hat{m}}
\def\ov{\overline}
\def\etapp{{\eta^{(')}}}
\def\fp{{f_{\eta'}^{(\bar cc)}}}
\def\half{{{1\over 2}}}
\def\pr{{\sl Phys. Rev.}~}
\def\prl{{\sl Phys. Rev. Lett.}~}
\def\pl{{\sl Phys. Lett.}~}
\def\np{{\sl Nucl. Phys.}~}
\def\zp{{\sl Z. Phys.}~}

\font\el=cmbx10 scaled \magstep2 {\obeylines \hfill NCKU-HEP-99-02
\hfill May, 1999}

\vskip 1.5 cm

\begin{center}
{\large {\bf Gauge Invariant and Infrared Finite Theory}}\par
{\large {\bf of Nonleptonic Heavy Meson Decays}}
\vskip 1.0cm
{\bf Hai-Yang Cheng}$^1$, {\bf Hsiang-nan Li}$^2$, and {\bf Kwei-Chou
Yang}$^1$
\vskip 0.5cm
$^1$Institute of Physics, Academia Sinica,\par
Taipei, Taiwan 105, Republic of China
\vskip 0.5cm
$^2$Department of Physics, National Cheng-Kung University, \par
Tainan, Taiwan 701, Republic of China
\end{center}
\vskip 1.0cm


\centerline{\bf Abstract}
\vskip 0.3cm
{\small
We show that the controversies on the gauge dependence and the infrared
singularity emerged in the generalized factorization approach for
nonleptonic heavy meson decays within the framework of the operator product
expansion can be resolved by perturbative QCD factorization theorem.
Gauge invariance of the decay amplitude is maintained under radiative
corrections by assuming on-shell external quarks. For on-shell external
quarks, infrared poles in radiative corrections have to be extracted using the
dimensional regularization. These poles, signifying nonperturbative dynamics
of a decay process, are absorbed into bound-state wave functions.
Various large logarithms produced in radiative corrections are summed
to all orders into the Wilson and Sudakov evolution factors. The remaining
finite part gives a hard subamplitude. A decay rate is
then factorized into a convolution of the hard subamplitude, the Wilson
coefficient, and the Sudakov factor with the
bound-state wave functions, all of which are well-defined and gauge
invariant.
}

\pagebreak


\section{Introduction}

The effective Hamiltonian is the standard starting point for describing
the nonleptonic weak decays of hadrons. Consider the decay $\ov B^0\to
D^+\pi^-$ as an example.
The relevant effective $\Delta B=1$ weak Hamiltonian is
\be
{\cal H}_{\rm eff} = {G_F\over\sqrt{2}}\, V_{cb}V_{ud}^*
\Big[c_1(\mu)O_1(\mu)+c_2(\mu)O_2(\mu)\Big],
\en
where
\be
O_1= (\bar cb)_\vma(\bar du)_\vma, \qquad\qquad  O_2 = (\bar
db)_\vma(\bar cu)_\vma,
\en
with $(\bar
q_1q_2)_{_{V\pm A}}\equiv\bar q_1\gamma_\mu(1\pm \gamma_5)q_2$. In
order to ensure the renormalization-scale and -scheme independence
for the physical amplitude, the matrix elements of 4-quark
operators have to be evaluated in the same renormalization scheme
as that for Wilson coefficients and renormalized at the same scale $\mu$.

Although the hadronic matrix element $\la O(\mu)\ra$ can be directly
calculated in the lattice framework, it is conventionally evaluated under
the factorization hypothesis so that $\la O(\mu)\ra$ is factorized into
the product of two matrix elements of
single currents, governed by decay constants and form factors.
In spite of its tremendous simplicity, the naive factorization approach
encounters two principal difficulties. First, it fails to describe
the color-suppressed weak decay modes. For example, the predicted decay
rate of $D^0\to\ov K^0\pi^0$ by naive factorization is too small by
two orders of magnitude compared
to experiment. Second, the hadronic matrix element under factorization is
renormalization scale $\mu$ independent as the vector or axial-vector
current is partially conserved. Consequently, the amplitude $c_i(\mu)
\la O\ra_{\rm fact}$ is not truly physical as the scale dependence of
Wilson coefficients does not get compensation from the matrix elements.

A plausible solution to the aforementioned scale problem is to
extract the $\mu$ dependence from the matrix element $\la
O(\mu)\ra$, and combine it with the $\mu$-dependent Wilson
coefficients to form $\mu$-independent effective Wilson
coefficients. After making a physical amplitude explicitly
$\mu$-independent, the factorization hypothesis is applied to the
hadronic matrix elements. However, the $\mu$-evolution factor
extracted from $\la O(\mu)\ra$ depends on an infrared cutoff,
which is originally implicit in $\la O(\mu)\ra$. Since an
off-shell external quark momentum is usually chosen as the
infrared cutoff, the $\mu$-evolution factor also contains a gauge
dependent term accompanied off-shell external quarks. Therefore,
this solution, though removes the scale and scheme dependence of a
physical amplitude in the framework of the factorization
hypothesis, often introduces the infrared cutoff and gauge
dependence.

In this paper we shall show that the above controversies can be
resolved by perturbative QCD (PQCD) factorization theorem. In this
formalism, partons, {\it i.e.}, external quarks, are assumed to be
on shell, and both ultraviolet and infrared divergences in
radiative corrections are isolated using the dimensional
regularization. Because external quarks are on shell, gauge
invariance of the decay amplitude is maintained under radiative
corrections. The obtained ultraviolet poles are subtracted in a
renormalization scheme, while the infrared poles are absorbed into
nonperturbative bound-state wave functions. Various large logarithms
produced in radiative corrections are summed to all orders into the Wilson
and Sudakov evolution factors. The remaining finite
piece is grouped into a hard decay subamplitude. The decay rate is
then factorized into the convolution of the hard subamplitude, the
Wilson coefficient, and the Sudakov factor with
the bound-state wave functions, all of which are well-defined and
gauge invariant. The partition of the nonperturbative and
perturbative contributions is quite arbitrary. Different
partitions correspond to different factorization schemes. However,
the decay rate, as the convolution of the above factors, is
independent of factorization schemes as it should be.

In Sec. II we review the conventional solutions to the scale and
scheme dependence present in the factorization hypothesis, and
their problems. Gauge invariance of radiative corrections is explicitly
justified to all orders in Sec. III. The PQCD approach is introduced in
Sec. IV. Explicit calculations of the evolution factor
$g_1(\mu)$ to be defined below are shown in Sec.~V. Section VI is the
conclusion.

\section{Gauge Dependence and Infrared Singularity}

The aforementioned scale problem with naive
factorization can be circumvented in two different approaches. In
the first approach, one incorporates nonfactorizable effects into the
effective coefficients \cite{Cheng94,Cheng96,Soares}:
\be
a_1^{\rm eff} = c_1(\mu) + c_2(\mu) \left({1\over N_c}
+\chi_1(\mu)\right)\,, \qquad \quad
a_2^{\rm eff} = c_2(\mu) + c_1(\mu)\left({1\over N_c} + \chi_2(\mu)\right)\,,
\en
where nonfactorizable terms are characterized by the parameters $\chi_i$.
For the decay $\ov B^0 \to D^+ \pi^-$, $\chi_1$ is given by
\be
c_2(\mu) \chi_1(\mu)= \left(c_1(\mu)+{c_2(\mu)\over N_c}\right)\vp^{(BD,
\pi)}_1(\mu)+ c_2(\mu) \vp_8^{(BD,\pi)}(\mu)\,,
\en
where
\be
\vp^{(BD,\pi)}_1&&=\frac{\langle D^+ \pi^-| (\bar cb)_\vma(\bar du)_\vma |
\ov B^0\rangle} {\langle D^+| (\bar cb)_\vma | \ov B^0 \rangle \langle
\pi^-|(\bar du)_\vma|0\rangle} -1\nonumber\,, \\
\vp^{(BD,\pi)}_8&&=\frac{\langle
D^+ \pi^-| {1\over 2} (\bar c \lambda^a b)_\vma (\bar d\lambda^a u)_\vma |
\ov B^0 \rangle} {\langle D^+| (\bar cb)_\vma | \ov B^0 \rangle
\langle \pi^-|(\bar du)_\vma|0\rangle}\,,
\en
are nonfactorizable terms originated from color-singlet and color-octet
currents, respectively, $(\bar q_1\lambda^a q_2)_\vma\equiv \bar
q_1\lambda^a \gamma_\mu(1-\gamma_5)q_2$.
The $\mu$ dependence of Wilson coefficients is assumed to be exactly
compensated by that of $\chi_i(\mu)$ \cite{Neubert}. That is, the
correct $\mu$ dependence
of the matrix elements is restored by the nonfactorized parameters
$\chi_i(\mu)$. However, there are two potential problems with this
approach. First, the renormalized 4-quark operator by itself
still depends on $\mu$, though the scale dependence of $\la O(\mu)\ra$
is lost in the factorization approximation. Second, to
the next-to-leading order (NLO), the Wilson coefficients depend on the
choice of the renormalization scheme. It is not clear if $\chi_i(\mu)$
can restore the scheme dependence of the matrix element.

In the second approach, it is postulated that $\la O(\mu)\ra$ is
related to the tree-level hadronic matrix element via the relation
$\la O(\mu)\ra=g(\mu)\la O\ra_{\rm tree}$ and that $g(\mu)$ is
independent of the external hadron states. Then schematically we
can write
\be
c(\mu)\la O(\mu)\ra=c(\mu)g(\mu)\la O\ra_{\rm tree}\equiv c^{\rm eff}\la O\ra
_{\rm tree}.
\label{ude}
\en
The factorization approximation is applied afterwards to the hadronic matrix
element of the operator $O$ at the tree level.
Since the tree-level matrix element $\la O\ra_{\rm tree}$ is renormalization
scheme and scale independent, so are the effective Wilson coefficients
$c_i^{\rm eff}$ and the effective parameters $a_i^{\rm eff}$ expressed by
\cite{Ali,CT98}
\begin{eqnarray} \label{aeff}
a_1^{\rm eff} = c_1^{\rm eff} + c_2^{\rm eff} \left({1\over N_c}
+\chi_1\right)\,, \qquad \quad
a_2^{\rm eff} = c_2^{\rm eff} + c_1^{\rm eff}\left({1\over N_c} +
\chi_2\right)\,.
\end{eqnarray}
However, the problem is that we do not know how to carry out
first-principles calculations of $\la O(\mu)\ra$ and hence
$g(\mu)$. It is natural to ask the question: Can $g(\mu)$ be
calculated at the quark level in the same way as the Wilson
coefficient $c(\mu)$ ? One of the salient features of the operator
product expansion (OPE) is that the determination of the
short-distance $c(\mu)$ is independent of the choice of external
states. Consequently, we can choose quarks as  external states in
order to extract $c(\mu)$. For simplicity, we consider a single
multiplicatively renormalizable 4-quark operator $O$ (say, $O_+$
or $O_-$) and assume massless quarks. The QCD-corrected weak
amplitude induced by $O$ in full theory is \be \label{full} A_{\rm
full}=\left[1+{\alpha_s\over 4\pi}\left(-{\gamma\over
2}\,\ln{M_W^2 \over -p^2}+a \right) \right]\la O\ra_q,
\en
where $\gamma$ is an anomalous dimension, $p$ is an off-shell momentum
of the external quark lines, which is introduced as an infrared cutoff,
and the non-logarithmic constant term $a$ in general depends on the
gauge chosen
for the gluon propagator. The subscript $q$ in (\ref{full}) emphasizes the
fact that the matrix element is evaluated between external quark states.
In effective theory, the renormalized $\la O(\mu)\ra_q$ is related to
$\la O\ra_q$ in full theory via
\be
\la O(\mu)\ra_q &=& \left[1+{\alpha_s\over 4\pi}\left(-{\gamma\over 2}\,\ln{
\mu^2\over -p^2}+r \right) \right]\la O\ra_q   \non \\
&\equiv & g'(\mu,-p^2,\lambda)\la O\ra_q,
\label{ou}
\en
where $g'$ indicates the perturbative corrections to the 4-quark
operator renormalized at the scale $\mu$. The constant term $r$ is
in general renormalization scheme, gauge and external momentum
dependent, which has the general expression \cite{Buras98}:
\be
r=r^{\rm NDR,HV}+\lambda r^\lambda,
\en
where NDR and HV stand for the naive dimension regularization and
't Hooft-Veltman renormalization schemes, respectively, and
$\lambda$ is a gauge parameter with $\lambda=0$ corresponding to
Landau gauge. Matching the effective theory with full theory,
$A_{\rm full}=A_{\rm eff}=c(\mu)\la O(\mu)\ra_q$, leads to
\be
c(\mu)=\,1+{\alpha_s\over 4\pi}\left(-{\gamma\over 2}\,\ln{M_W^2
\over \mu^2}+d \right),
\en
where $d=a-r$. Evidently, the Wilson coefficient is independent of the
infrared cutoff and it is gauge invariant as the gauge dependence is
compensated between $a$ and $r$.
Of course, $c(\mu)$ is still
renormalization scheme and scale dependent.

Since $A_{\rm eff}$ in full theory [Eq.~(8)] is $\mu$ and scheme
independent, it is obvious that
\be
c'^{\rm eff}=c(\mu)g'(\mu,-p^2,\lambda)
\label{gfc}
\en
is also independent of the choice of the scheme and scale.
Unfortunately, $c'^{\rm eff}$ is subject to the ambiguities of the
infrared cutoff and gauge dependence, which come along with $g'$ extracted
from $\la O(\mu)\ra_q$. As stressed in
\cite{Buras98}, the gauge and infrared dependence always appears
as long as the matrix elements of operators are calculated between
quark states. Therefore, it is unreliable to define the effective
Wilson coefficients by applying the existing calculations in the literature.
The reason
has been implicitly pointed out in ~\cite{ACMP} that ``off-shell
renormalized vertices of gauge-invariant operators are in general
gauge dependent".

The existing problems associated with the off-shell regularization
scheme are as follows:

1. When working with off-shell fermions, there exists the
so-called $P$ operator~\cite{ACMP} e.g., ${\slash\!\!\!
p}(1-\gamma_5) \otimes {\slash\!\!\! p} (1-\gamma_5)$ which cannot
be removed by the equation of motion.


2. The finite terms are external momentum dependent (see Fig.~3 of
\cite{CFMR}) and they are obtained in some specific condition. For
example, two incoming fermion legs 1,2 and two outgoing legs 3,4
with external momentum $p$ are chosen in Figs. 3a and 3c of
\cite{CFMR,Ciuchini}, while legs 1,3 incoming and 2,4 outgoing
with $p$ in Fig. 3b.

Hence we cannot avoid the gauge problems if adopting off-shell
fermions and the finite parts of $c'^{\rm eff}$ are not
well-defined. To circumvent this difficulty, we should work in a
physical on-shell scheme and employ the dimensional regularization
for infrared divergences. Gauge invariance of the decay amplitude
is maintained under radiative corrections, and the infrared poles
are absorbed into the hadronic matrix element as stated in the
Introduction. Consequently, the effective coefficient $c^{\rm
eff}=c(\mu)g(\mu)$ does not suffer from the gauge ambiguity.

\section{GAUGE INVARIANCE IN ON-SHELL REGULARIZATION}

In this section we show that gluon exchanges among the on-shell quarks
involved in heavy meson decays, including the spectator quarks, indeed give
gauge invariant contributions. We present the proof in the covariant gauge
$\partial\cdot A=0$, in which the gluon propagator is given by
$(-i/l^2)N^{\mu\nu}(l)$ with
\begin{equation}
N^{\mu\nu}(l)=g^{\mu\nu}-\left(1-{\lambda}\right)
\frac{l^\mu l^\nu}{l^2}\;,
\label{gp}
\end{equation}
where $\lambda$ is the gauge parameter. We shall show that the
quark amplitude $A_{\rm full}$ with the spectator quarks included
are independent of $\lambda$ to all orders, namely,
\begin{equation}
{\lambda}\frac{d A_{\rm full}}{d\lambda}=0\;.
\label{diff}
\end{equation}
The differential operator applies only to gluon propagators,
leading to
\begin{eqnarray}
{\lambda}\frac{d}{d\lambda}N^{\mu\nu}=
{\lambda}\frac{l^\mu l^\nu}{l^2}
=v_\alpha[l^\mu N^{\alpha\nu}+l^\nu N^{\alpha\mu}]\;,
\label{digp}
\end{eqnarray}
with the special vertex $v_\alpha=l_\alpha/(2l^2)$. The loop
momentum $l^\mu$ ($l^\nu$) carried by the differentiated gluon
contracts with a vertex in $A_{\rm full}$, which is then replaced
by the special vertex $v_\alpha$. Eq. (\ref{digp}) is graphically
described by the first expression in Fig.~1, where the arrrow
represents $l^\mu$ ($l^\nu$) contracting with the gluon vertex,
and the square represents $v_\alpha$.

\begin{figure}[tb]
\psfig{figure=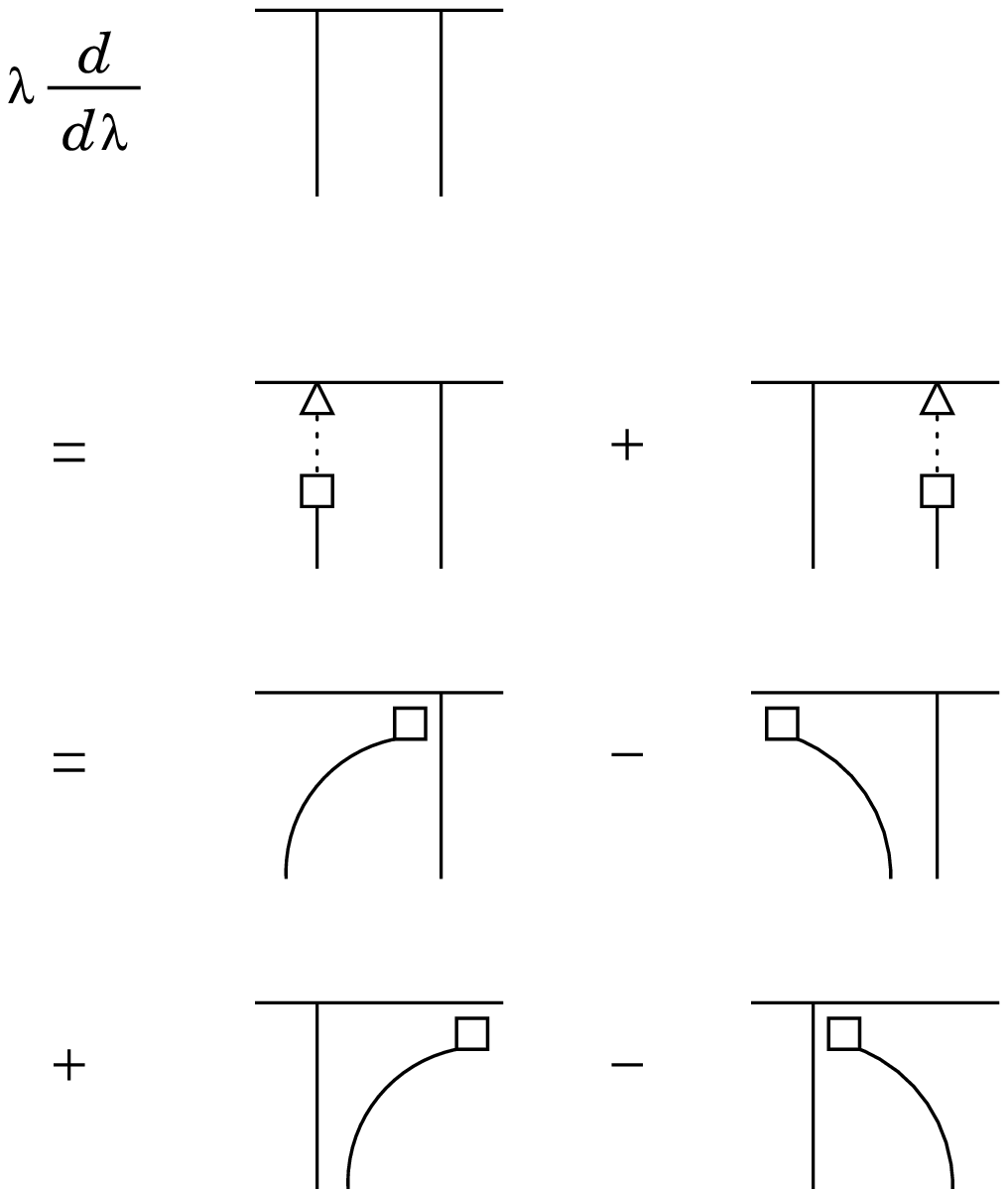,height=3.5in} \vspace{0.4cm}
    \caption{{\small Graphical representation for the proof of
Eq.~(\ref{diff}).}}
\end{figure}

The contraction of $l^\mu$ ($l^\nu$) leads to the Ward identity \cite{L1}
shown in the second expression of Fig.~1, where the solid lines may
represent quarks or gluons.
Summing all the diagrams with various differentiated gluons, those
embedding the special vertices cancel by pairs. For example, the
pair cancellation occurs between the first and last diagrams in the second
expression of Fig.~1. Only the diagram, in which the special vertex moves to
the outer end of the quark line, is left. This diagram comes
from the second term in the following expression,
\begin{equation}\label{eoq}
\frac{i(\not k+\not l+M)}{(k+l)^2-M^2}(-i\not l)u(k)
=u(k)-\frac{\not k+\not l+M}{(k+l)^2-M^2}(\not k-M) u(k)\;,
\end{equation}
where $u$ is the fermion spinor associated with an external quark.
The first term is canceled by the term from the contraction of $l$
with the adjacent vertex. If all the external quarks are on shell,
the second term vanishes because of the equation of motion $(\not
k-M) u(k)=0$. Then we arrive at the desired result (\ref{diff}).

We take one-loop corrections as an example to elucidate the above
proof. We consider only the gauge-dependent part of the gluon
propagator [see Eq.~(\ref{gp})],
\begin{equation}
\frac{-i}{l^2}\left[-\left(1-{\lambda}\right)
\frac{l^\mu l^\nu}{l^2}\right]\;, \label{gpgauge}
\end{equation}
in the loop calculations and demonstrate that the result vanishes
after summing all the diagrams. The gauge-dependent part of
Fig.~2(a) reads
\begin{eqnarray}\label{Ia}
&&I_a^{gauge}= i{ G_F \over\sqrt{2}} V_{\rm
CKM}\biggl(1-{\lambda}\biggr)
 g_s^2 \mu^\epsilon\nonumber\\
&&\times \int{d^D k\over (2\pi)^D} \frac{\bar u_3 t^a\gamma_\mu
(1-\gamma_5)(\not
 p_1+\not k+M_1) \not k u_1 \; \bar u_4 t^a \gamma^\mu(1-\gamma_5)(\not
 p_2-\not k+M_2)\not k u_2}{k^4[(k+p_1)^2-M_1^2][(k-p_2)^2-M_2^2]}\,,
\end{eqnarray}
where $V_{\rm CKM}$ is the relevant Cabibbo-Kobayashi-Maskawa
matrix elements.

\begin{figure}[tb]
\psfig{figure=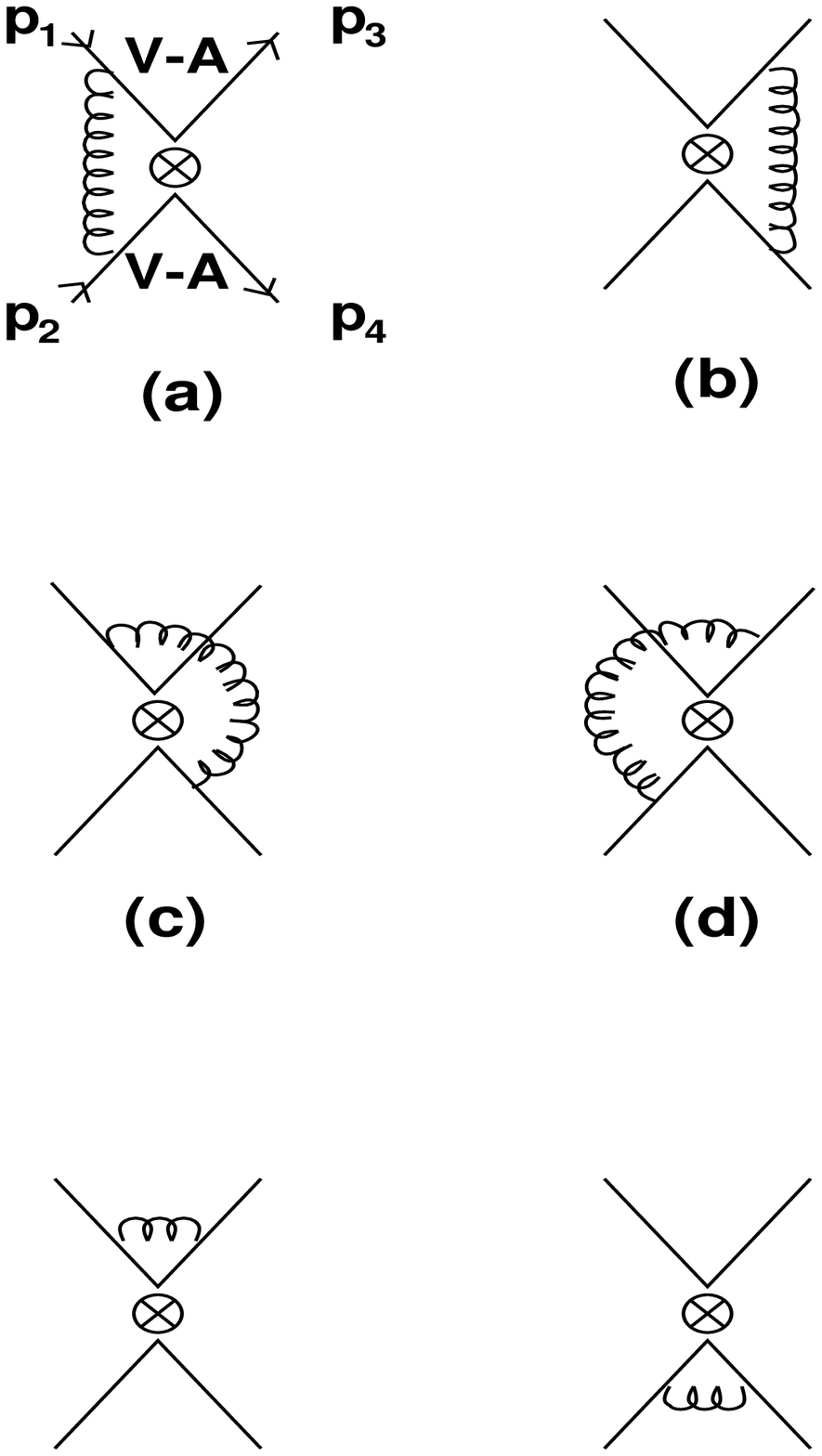,height=3.5in} \vspace{0.9cm}
    \caption{{\small
     Vertex corrections to the 4-quark operators $O_1$ and $O_2$.}}
     \label{fig:4quark}
\end{figure}

To proceed, we replace the $\not k$ which is adjacent to $u_1$ as
\begin{eqnarray}
\not k=\not p_1 +\not k -M_1 -(\not p_1 -M_1)\,,
\end{eqnarray}
and the $\not k$ adjacent to $u_2$ as
\begin{eqnarray}
\not k=-\not p_2 +\not k +M_2 +(\not p_2 -M_2)\,.
\end{eqnarray}
Applying Eq.~(\ref{eoq}) and the equation of motion, $(\not p-M)
u=0$, Eq.~(\ref{Ia}) becomes
\begin{eqnarray}\label{Ianew}
&&I_a^{gauge}= i{ G_F \over\sqrt{2}} V_{\rm
CKM}\biggl(1-{\lambda}\biggr)
 g_s^2 \mu^\epsilon \int{d^D k\over (2\pi)^D} \frac{\bar u_3 t^a\gamma_\mu
(1-\gamma_5) u_1 \; \bar u_4 t^a\gamma^\mu(1-\gamma_5)u_2}{k^4}\,.
\end{eqnarray}
Likewise, one can apply the same trick to the calculations of
Figs.~2(b)-(f) and obtain
$I_b^{gauge}=-I_c^{gauge}=-I_d^{gauge}=I_a^{gauge}$, and
\begin{eqnarray}\label{Iefnew}
&&I_e^{gauge}=I_f^{gauge}\nonumber\\
&&= i{ G_F \over\sqrt{2}}
V_{\rm CKM}\biggl(1-{\lambda}\biggr)
 g_s^2 \mu^\epsilon C_F\int{d^D k\over (2\pi)^D} \frac{1}{k^4}{\bar u_3
 \gamma_\mu
(1-\gamma_5) u_1 \; \bar u_4 \gamma^\mu(1-\gamma_5)u_2}\nonumber\\
&&= -{ G_F \over\sqrt{2}} V_{\rm
CKM}\biggl(1-{\lambda}\biggr)
 {\alpha_s\over 4\pi} C_F\biggl({2\over \epsilon_{_{\rm UV}}}-{2\over
 \epsilon_{_{\rm IR}}}
 \biggr) {\bar u_3 \gamma_\mu
(1-\gamma_5) u_1 \; \bar u_4 \gamma^\mu(1-\gamma_5)u_2}\,,
\end{eqnarray}
where $\epsilon_{_{\rm UV(IR)}}=4-D$ is the ultraviolet (infrared)
pole, and $C_F=(N_c^2-1)/(2N_c)$ with $N_c$ being the number of
colors. After lengthy but straightforward calculation, the
renormalization constant of a fermion with mass $M$ is found to be
\begin{eqnarray}
Z_2 &=& 1-{\alpha_s\over 4\pi} C_F\biggl({2\over \epsilon_{_{\rm
UV}}}-3\gamma_E+4-3\ln{M^2\over 4\pi^2\mu^2}-{4\over
\epsilon_{_{\rm IR}}}\biggr)   \non\\ &&+(1-\lambda){\alpha_s\over
4\pi}C_F\biggl({2\over \epsilon_{_{\rm UV}}}-{2\over
\epsilon_{_{\rm IR}}}\biggr).
\end{eqnarray}
We see that, contrary to the gauge-independent part of $Z_2-1$,
the gauge-dependent contribution due to the fermion wavefunction
renormalization
\begin{eqnarray}
 { G_F \over\sqrt{2}} V_{\rm
CKM}\biggl(1-{\lambda}\biggr)
 {\alpha_s\over 4\pi} C_F\biggl({2\over \epsilon_{_{\rm UV}}}-{2\over
 \epsilon_{_{\rm IR}}}\biggr)
 {\bar u_3 \gamma_\mu
(1-\gamma_5) u_1 \; \bar u_4 \gamma^\mu(1-\gamma_5)u_2}\,
\end{eqnarray}
is free of mass singularity. Summing over all the contributions,
it is obvious that the final result indeed vanishes as it should
be.

\section{PQCD FACTORIZATION THEOREM}

We have shown that radiative corrections to a decay amplitude of
on-shell external quarks are gauge invariant to all orders. The
one-loop diagrams have been evaluated explicitly, whose results
confirm our proof. Next we shall explain how to treat the infrared
poles in the PQCD factorization theorem. The one-loop
contributions in full theory are ultraviolet finite because of the
current conservation stated above. The existence of the infrared poles
simply signifies the nonperturbative dynamics, which demands the inclusion
of bound-state effects into the formalism of heavy meson decays. The
standard treatment of infrared poles is to absorb them into a universal
meson wave function. To absorb the infrared poles associated with the $b$
quark, such as those from the self-energy corrections, it is necessary to
introduce a $B$ meson wave function. That is, we must take into
account the spectator quark of the $B$ meson in order to develop a
complete theory of heavy meson decays.

Accordingly, the decay amplitude $A_{\rm full}$ to one-loop in full theory
can be rewritten as
\begin{eqnarray}
A_{\rm
full}&=&1+\frac{\alpha_s}{4\pi}\left(\frac{c}{\epsilon_{IR}}
+\gamma\ln\frac{M_W}{M_b}+\gamma'\ln\frac{M_b}{\mu_f}+a\right)
\nonumber\\
&=&\left[1+\frac{\alpha_s}{4\pi}\left(\gamma\ln\frac{M_W}{M_b}
+\gamma'\ln\frac{M_b}{\mu_f}+a\right)\right]
\left(1+\frac{\alpha_s}{4\pi}\frac{c}{\epsilon_{_{\rm IR}}}\right)
+O(\alpha_s^2)\;,
\end{eqnarray}
where the factorization scale $\mu_f$ arises from the dimensional
regularization of infrared divergences, and the factorization of
the infrared pole is performed in the minimal subtraction scheme.
The anomalous dimensions of the logarithms $\ln(M_W/M_b)$ and
$\ln(M_b/\mu_f)$, $\gamma$ and $\gamma'$, respectively, are
different, since the latter involves an extra contribution related
to the spectator quark. The factor containing the infrared pole
can be formulated as a matrix element of a nonlocal operator,
which is the definition of a meson wave function $\phi(\mu_f)$. A
wave function, describing the amplitude that a parton carries a
fraction of the meson momentum, cannot be derived in perturbation
theory. It must be parametrized as a function of parton momentum
fraction.

We further factorize the infrared finite part into
\begin{eqnarray}
A_{\rm
full}&=&\left[1+\frac{\alpha_s}{4\pi}\left(\gamma\ln\frac{M_W}{\mu}+
a'\right)\right]\left[1+\frac{\alpha_s}{4\pi}\left(\gamma\ln\frac{\mu}{M_b}
+\gamma'\ln\frac{M_b}{\mu_f}+a-a'\right)\right] \nonumber\\ &
&\times \left(1+\frac{\alpha_s}{4\pi}\frac{c}{\epsilon_{_{\rm
IR}}}\right) +O(\alpha_s^2)\;.
\end{eqnarray}
The first factor, characterized by the matching scale $M_W$, is identified as
the Wilson coefficient $c(\mu)$ after summing $\ln(M_W/\mu)$ to all orders
using renormalization group (RG) equations. The second factor,
characterized by the $b$ quark mass $M_b$, is the hard subamplitude
which will be denoted by $H(M_b,\mu,\mu_f)$ below.
Extending the above procedures to all orders, we obtain the factorization
formula for $B$ meson (not $b$ quark) decays,
\begin{eqnarray}
A_{\rm full}=c(\mu)H(M_b,\mu,\mu_f)\phi(\mu_f)\;,
\label{three}
\end{eqnarray}
which is exactly the three-scale factorization formula for exclusive
nonleptonic decays derived in \cite{CL}. Note that
Eq.~(\ref{three}) in fact denotes a convolution relation, because the
momentum fractions should be integrated out.

Compared to Eq.~(\ref{ude}), the matrix element
$\langle O(\mu)\rangle$ corresponds to
\begin{eqnarray}
\langle O(\mu)\rangle=H(M_b,\mu,\mu_f)\phi(\mu_f)\;.
\end{eqnarray}
Summing $\ln(\mu/M_b)$ in $H$ to all orders using RG equations, we obtain
an evolution factor $g_1(\mu)$, whose behavior from $\mu$ to $M_b$ is
governed by the same anomalous dimension as that of $c(\mu)$. Summing
$\ln(M_b/\mu_f)$ in $H$ to all orders, we obtain another factor
$g_2(\mu_f)$ describing the evolution from $M_b$ to $\mu_f$, whose
anomalous dimension differs from that of $c(\mu)$
because of the inclusion of the dynamics associated with spectator quarks.
Note that $g_2$ is part of the Sudakov evolution obtained in \cite{CL}.
Hence, the $\mu$ dependence of $H$ is extracted as
\begin{equation}
H(M_b,\mu,\mu_f)=g_1(\mu)g_2(\mu_f)H(M_b,M_b,M_b)\;.
\label{mrg}
\end{equation}
The combination of $c$, $g_1$, and $g_2$ leads to the effective coefficient
\begin{eqnarray}
c^{\rm eff}=c(\mu)g_1(\mu)g_2(\mu_f)\;,
\label{nef}
\end{eqnarray}
which is not only $\mu$ and scheme independent but also gauge invariant.
The factor $g(\mu)$ in Eq.~(\ref{ude}) can be identified as
$g_1(\mu)g_2(\mu_f)$, which describes the evolution down to the factorization
scale. However, $g_1(\mu)g_2(\mu_f)$ contains a matching condition at the
scale $M_b$ between the Wilson and Sudakov evolutions with different
anomalous dimensions. Therefore, there is an ambiguity of the matching
condition: the two evolutions can also match at $rM_b$ with $r$ a constant
of order unity. Obviously, Eq.~(\ref{nef}) is subtler than the naive
definition of $c^{\rm eff}$ in Eq.~(\ref{ude}).

The matrix element $\langle O\rangle_{\rm tree}$ in Eq.~(\ref{ude}) is
then identified as
\begin{eqnarray}
\langle O\rangle_{\rm tree}=H_{\rm tree}(M_b,M_b,M_b)\phi(\mu_f)\;,
\label{otr}
\end{eqnarray}
where the hard subamplitude is evaluated to lowest order with one
hard gluon exchange, since all large logarithms have been
organized by RG equations. We emphasize that the factorization
hypothesis for $\langle O\rangle_{\rm tree}$ in the conventional
approach is not necessary in the PQCD formalism. The purpose of
the factorization hypothesis is to simplify the decay amplitude
into products of decay constants and form factors, which are then
parametrized as various models. To have a better fit to
experimental data, nonfactorizable contributions, parametrized as
$\chi$ [see Eq.~(7)], are included. Note that $H_{\rm tree}$ in
the PQCD approach includes both factorizable contributions (form
factors), when the hard gluon attaches to the two quarks in a
meson, and nonfactorizable contributions (octet amplitudes), when
the hard gluon attaches to the quarks in different mesons.
Therefore, we may compute all possible diagrams for $H_{\rm tree}$
\cite {YL} and convolute them with the same meson wave functions
$\phi$. That is, we use the single parametrization, {\it i.e.},
the meson wave functions, for both factorizable and
nonfactorizable contributions based on Eq.~(\ref{otr}). In this
sense the PQCD formalism is more systematic.

At last, we explain how to handle the nonperturbative meson wave
functions with the dependence of the factorization scale $\mu_f$.
It can be shown that these wave functions are universal for all
decay processes involving the same mesons. For example, the $B$
meson wave function for the nonleptonic decays $B\to
D^{(*)}\pi(\rho)$ and for the radiative decay $B\to K^*\gamma$ is
the same. This universality can be easily understood, since a wave
function collects long-distance (infrared) dynamics, which should
be insensitive to short-distance dynamics involved in the decay of
the $b$ quark into light quarks with large energy release. Based
on the universality of wave functions, the application of
factorization formulas is as follows \cite{LM}. We evaluate the
Wilson and Sudakov evolutions down to a factorization scale
$\mu_f$ and the hard subamplitude for a decay mode, say, $B\to
K^*\gamma$, in perturbation theory. These calculations are simply
performed at the quark level with infrared poles dropped (in the
minimal subtraction scheme). Adjust the $B$ meson wave function
such that the predictions from the relevant factorization formula
match the experimental data. At this stage, we determine the $B$
meson wave function defined at the scale $\mu_f$. Then evaluate
the Wilson and Sudakov evolutions down to the same scale $\mu_f$
and the hard subamplitude for another decay, say, $B\to D\pi$.
Convolute them with the same $B$ meson wave function and make
predictions. At this stage, there are no free parameters in the formalism.
With the above strategy, the PQCD factorization
theorem possesses predictive power.

The main uncertainties in the PQCD factorization theorem come from
higher-order corrections to the hard subamplitude and higher-twist
corrections from the Fock states other than the leading one with
only valence quarks which we are considering here. According to
Eq.~(\ref{mrg}), the argument of the running coupling constant in
the hard subamplitude $H(M_b,M_b,M_b)$ should be set to the $b$
quark mass $M_b$, implying that the next-to-leading-order diagrams
give about $\alpha_s(M_b)/\pi\sim 10\%$ corrections. Since $H$ is
characterized by $M_b$, the next-to-leading-twist correction from
the Fock state with one more parton entering the hard subamplitude
is about $\mu_f/M_b\sim 10\%$. Note that meson wave functions are
usually defined at the factorization scale $\mu_f\sim 0.5$ GeV
\cite{CZ}. We believe that other nonperturbative corrections, such
as final-state interactions, should play a minor role because of
the large energy release involved in two-body $B$ meson decays. If
the hadronic matrix elements are evaluated using the factorization
approximation (i.e. vacuum insertion approximation), the related
uncertainties have been discussed in length in \cite{CCTY}.

In conclusion, all the factors in the PQCD formalism are
well-defined (including the nonperturbative meson wave functions)
and gauge invariant. Physical quantities obtained in this
formalism are scale and scheme independent. We have applied this
approach to exclusive semileptonic, nonleptonic, and radiative $B$
meson decays and the results are very successful. Nonfactorizable
contributions have been calculated, and found to play an important
role in the decays $B\to J/\psi K^{(*)}$ \cite{YL}. The opposite
signs of $a_2/a_1$ in bottom and charm decays have been explained
by the effects of the Wilson evolution \cite{YL}. The mechanism
for the sign change of the nonfactorizable contributions in bottom
and charm decays have also been explored \cite{LT}, which is
closely related to the success and failure of the large-$N_c$
limit in charm and bottom decays, respectively.

\section{effective Wilson coefficients}

In this section we present the results for the evolution factor
$g_1(\mu)$ which describes the evolution from the scale $\mu$ to
$M_b$ for the current-current operators $O_1$ and
$O_2$.\footnote{The complete results of $g_1(\mu)$ for $\Delta
B=1$ transition current-current operators $O_1,O_2$, QCD-penguin
operators $O_3,\cdots,O_6$ and electroweak penguin operators
$O_7,\cdots,O_{10}$ are given in \cite{CCTY}.}  Setting
$\mu_f=M_b$, the effective Wilson coefficients obtained from the
one-loop vertex diagrams Figs. 2(a)-(f) for the operators $O_i$
have the form:
\be
c_1^{\rm eff}\Big|_{\mu_f=M_b} &=& c_1(\mu)+{\alpha_s\over
4\pi}\left(\gamma^{(0)T}\ln{M_b\over \mu}+r^T\right)_{1i}c_i(\mu),
\non
\\ c_2^{\rm eff}\Big|_{\mu_f=M_b} &=& c_2(\mu)+{\alpha_s\over
4\pi}\left(\gamma^{(0)T}\ln{M_b\over \mu}+r^T\right)_{2i}c_i(\mu),
\en
where the superscript $T$ denotes a transpose of the matrix, and
the anomalous dimension matrix $\gamma^{(0)}$ due to the one-loop
vertex corrections has the well-known expression
\be
\label{gamma} \gamma^{(0)}=\left(\matrix{ -2 & 6 \cr 6 & -2
\cr}\right).
\en
The matrix $r$ gives momentum-independent constant terms which
depend on the treatment of $\gamma_5$. Working in the (massless)
on-shell scheme and assuming zero momentum transfer squared
between color-singlet currents, i.e. $(p_1-p_3)^2=0$ as well as
$(p_1+p_2)^2=(-p_2+p_3)^2\approx m_b^2$ for $O_1$ operators and
$(p_1-p_4)^2=0$, $(p_1+p_2)^2=(-p_2+p_4)^2\approx m_b^2$ for $O_2$
operators (see Fig. \ref{fig:4quark} for momentum notation), we
obtain
\be
\label{rind} r_{\rm NDR}=\left(\matrix{ 3 & -9 \cr -9 & 3
\cr}\right), \qquad\qquad r_{\rm HV}=\left(\matrix{ {7\over 3} &
-7 \cr -7 & {7\over 3} \cr}\right)
\en
in NDR and HV schemes, respectively. It should be accentuated
that, contrary to the previous work \cite{Ali}
\be
\label{rndr} r_{\rm NDR}^{\rm \lambda=0}=\left(\matrix{ {7\over 3}
& -7 \cr -7 & {7\over 3} \cr}\right), \qquad\qquad r_{\rm
HV}^{\lambda=0}=\left(\matrix{ 7 & -5 \cr -5 & 7 \cr}\right),
\en
obtained in Landau gauge and off-shell regularization, the matrix
$r$ given in (\ref{rind}) is gauge invariant !

Two remarks are in order. First, there are infrared double poles,
i.e., $1/\epsilon_{_{\rm IR}}^2$, in the amplitudes of Figs.
2(a)-2(d), but they are canceled out when adding all amplitudes
together. Second, care must be taken when applying the projection
method to reduce the tensor products of Dirac matrices to the form
$\Gamma\otimes\Gamma$ with $\Gamma=\gamma_\mu(1-\gamma_5)$.  For
example, a direct evaluation of the tensor product
$\gamma_\alpha\not p_1\Gamma\otimes\Gamma\not p_2\gamma^\alpha$
yields $(\epsilon=4-D$)
\be
\label{pro1} \gamma_\alpha\not p_1\Gamma\otimes\Gamma\not
p_2\gamma^\alpha=-\epsilon(p_1\cdot p_2)\Gamma\otimes\Gamma
\en
in the NDR scheme with the on-shell condition being applied first
to the massless quarks followed by Fierz transformation, whereas
the projection method of \cite{Buras90,Buras92,CFMR} leads to
\be
\label{pro2} \gamma_\alpha\not p_1\Gamma\otimes\Gamma\not
p_2\gamma^\alpha=(p_1\cdot p_2)\Gamma\otimes\Gamma+E,
\en
where $E$ stands for the evanescent operator (EO). This means that
it is incorrect to take the coefficient of $\Gamma\otimes\Gamma$
in Eq. (\ref{pro2}) directly without taking into account the
effect of EOs. Note that we have applied Eq. (\ref{pro1}) to show
the absence of infrared double poles in the total amplitude.

In order to check the scheme and scale independence of $c^{\rm
eff}_i$, it is convenient to work in the diagonal basis in which
the operators $O_\pm={1\over 2}(O_1\pm O_2)$ do not mix under
renormalization. Then (see e.g. \cite{Buras96} for the general
expression of $c(\mu)$)
\be
c_\pm^{\rm eff}\Big|_{\mu_f=M_b} &=& c_\pm(\mu)g_1^\pm (\mu)
\nonumber \\ &=&\left[1+{\alpha_s(\mu)\over
4\pi}(r^T_\pm+J_\pm)\right] \left[1+{\alpha_s(M_W)\over
4\pi}(B_\pm-J_\pm)\right] \non\\ &\times& \left(
\left[{\alpha_s(M_W)\over
\alpha_s(\mu)}\right]^{\gamma^{(0)}_\pm/(2\beta_0)}+{\alpha_s\over
4\pi}\,{\gamma_\pm^{(0)}\over 2}\ln{M_b^2\over \mu^2}\right),
\en
where $c_\pm=c_1\pm c_2$, $\beta_0=11-{2\over 3}n_f$ with $n_f$
being the number of flavors between $M_W$ and $\mu$ scales,
$B_\pm$ specifies the initial condition of $c(m_W)$:
$c(m_W)=1+{\alpha_s(m_W)\over 4\pi}B_\pm$ and it is
$\gamma_5$-scheme dependent, and $J_\pm=\gamma^{(0)}_\pm
\beta_1/(2\beta_0^2)- \tilde\gamma^{(1)}_\pm /(2\beta_0)$, with
$\beta_1=102-38n_f/3$. The scheme-dependent anomalous dimensions
$\tilde\gamma^{(1)}_\pm$ are given by \cite{Buras96,Buras98}: \be
\tilde\gamma^{(1)}_\pm=\gamma^{(1)}_\pm-2\gamma_J={3\mp 1\over
6}(-21\pm {4\over 3}n_f-2\beta_0\kappa_\pm),
\en
where $\gamma^{(1)}_\pm$ are the two-loop anomalous dimensions of $O_\pm$,
 $\gamma_J$ is the anomalous dimension of the weak
current in full theory, and the parameter $\kappa_\pm$
distinguishes various renormalization schemes: $\kappa_\pm=0$ in
the NDR scheme and $\kappa_\pm=\mp 4$ in the HV scheme. As shown
in \cite{Buras96}, $B_\pm-J_\pm$ is $\gamma_5$-scheme independent.
Therefore, the effective Wilson coefficients $c_\pm^{\rm eff}$ are
scheme independent if we are able to show that $(r^T+J)$ is
independent of the choice of renormalization scheme. Since the
short-distance Wilson coefficients are independent of the choice
of external states, one can show the independence of
$\gamma^{(1)}_\pm-2\gamma_J$ from external states \cite{Buras96}.
In the on-shell scheme, $\gamma_J$ vanishes up to the two-loop
level. It follows from Eq. (\ref{rind}) that $r_+=-6$, $-7+7/3$
and $r_-=12$, $7+7/3$ in NDR and HV schemes, respectively, with
fermions being on-shell. Then it is easily seen that $ r^T+J$ is
indeed renormalization scheme independent. To the leading
logarithmic approximation,
\be
\left({\alpha_s(M_W)\over
\alpha_s(\mu)}\right)^{\gamma^{(0)}_\pm/(2\beta_0)}\approx
1-{\alpha_s\over 4\pi}\,{\gamma_\pm^{(0)}\over
2}\ln{M_W^2\over\mu^2}.
\en
Hence, the scale independence of the effective Wilson coefficients
follows.

Since the weak current is partially conserved, its anomalous
dimension $\gamma_J$ is zero. However, if fermions are off-shell,
$\gamma_J$ is non-vanishing at the two-loop level in the HV
scheme. To maintain the requirement that $\gamma_J=0$, one can
force a vanishing $\gamma_J$ in this case by applying a finite
renormalization term to the weak current. (Note that in this new
choice, $\gamma^{(1)'}_\pm=\gamma^{(1)}_\pm$,
$B'_\pm=B_\pm-\gamma_J/\beta_0$, and $B'_\pm -J'_\pm$ is still
scheme independent.) Using the identity
$\gamma^{(1)}$(on-shell)=$\gamma^{(1)}$(off-shell)$-2\gamma_J$, we
find that $\gamma^{(1)}$ in the off-shell fermion scheme is given
by \cite{Buras92}:
\be
\label{r1} \gamma^{(1)}_{\rm NDR} &=& \left(\matrix{ -{21\over
2}-{2\over 9}n_f & {7\over 2}+{2\over 3}n_f \cr {7\over 2}+{2\over
3}n_f & -{21\over 2}-{2\over 9}n_f  \cr}\right),  \non \\
\gamma^{(1)}_{\rm HV} &=& \left(\matrix{ {553\over 6}-{58\over
9}n_f & {95\over 2}-2n_f \cr {95\over 2}-2n_f & {553\over
6}-{58\over 9}n_f \cr}\right).
\en
From Eqs. (\ref{rndr}) and (\ref{r1}), it is straightforward to
show that $r_\pm +J_\pm$ is $\gamma_5$-scheme independent in the
off-shell regularization. As a result, $c^{\rm eff}$ is
renormalization scheme independent, irrespective of the fermion
state, on-shell or off-shell.

\section{CONCLUSION}

In this paper we have shown how to construct a gauge invariant and
infrared finite theory of exclusive nonleptonic $B$ meson decays
based on PQCD factorization theorem. Gauge invariance is
maintained under radiative corrections by working  in the physical
on-shell scheme. The infrared divergences in radiative corrections
should be then isolated using the dimensional regularization. The
resultant infrared poles are absorbed into the universal meson
wave functions, which can be determined once for all from
experimental data. The absorption of the poles associated with the
$b$ quark requires the inclusion of the spectator quark into the
theory. The remaining finite contributions form a hard
subamplitude. Applying RG analyses to sum various large logarithms
in the above factorization formula, the scale and scheme
dependences are removed. Hence, in the PQCD formalism physical
quantities are guaranteed to be gauge invariant, infrared finite,
scale and scheme independent. By working out the evolution factor
$g_1(\mu)$ explicitly, we have constructed gauge invariant, scale
and scheme independent effective Wilson coefficients
$c(\mu)g_1(\mu)$ at the factorization scale $\mu_f=M_b$. We have
shown explicitly that $c^{\rm eff}$ are renormalization scheme and
scale independent.

We shall take one of the exclusive nonleptonic $B$ meson decay
modes as an example to demonstrate how to construct a
factorization formula explicitly. This work will be published
elsewhere.

\vskip 1cm
\centerline{\bf Acknowledgement}
\vskip 0.4cm
This work was supported in part by the National Science Council of R.O.C.
under the Grant Nos. NSC-88-2112-M-006-013 and NSC-88-2112-M-001-006.


\end{document}